\documentclass[12pt,preprint]{aastex}

\begin{document}
\def\etal{{\sl et al.~}}
\def\Teff{{$\rm{T_{eff}}$}}
\def\dTeff{{$\langle \Delta \rm{T_{eff}} \rangle$}}

\title{Chemical Abundances in 35 Metal-Poor Stars. I. Basic Data}

\author{Jeong-Deok Lee\altaffilmark{1, 2},
        Sang-Gak Lee\altaffilmark{2},
        \& Kang-Min Kim\altaffilmark{3}}

\email{leejd@astro.snu.ac.kr}

\altaffiltext{1}{Astrophysics Research Center for the Structure
and Evolution of the Cosmos (ARCSEC), Sejong University, Seoul,
143-747, South Korea}

\altaffiltext{2}{Department of Physics and Astronomy, Seoul
National University, Seoul 151-742, South Korea}

\altaffiltext{3}{Korea Astronomy \& Space Science Institute, 61-1
Whaam-dong, Youseong-gu, Taejeon 305-348, South Korea}

\received{updated 1/Aug/2007}

\shorttitle{Chemical Abundances in Metal-Poor Stars}
\shortauthors{Lee, Lee, \& Kim }

\begin{abstract}
We carried out a homogeneous abundance study for various elements,
including $\alpha$-elements, iron peak elements and $n$-capture
elements for 35 metal-poor stars with a wide metallicity range
($-3.0\lesssim$[Fe/H]$\lesssim-0.5$). High-resolution
($R\simeq30$k), high signal-to-noise($S/N\geq110$) spectra with a
wavelength range of 3800 to 10500 \AA using the Bohyunsan Optical
Echelle Spectrograph (BOES). Equivalent widths were measured by
means of the Gaussian-fitting method for numerous isolated weak
lines of elements. Atmospheric parameters were determined by a
self-consistent LTE analysis technique using Fe I and Fe II lines.
In this study, we present the EWs of lines and atmospheric
parameters for 35 metal-poor stars.

\end{abstract}


\section{INTRODUCTION}

The abundance patterns of metal-poor stars provide information on
the formation, evolution, and supernova explosions of massive
stars formed in the early epoch of the Galaxy formation
\citep[e.g.][]{tinsley80, pagel97,matteu01}. They also provide
important clues for understanding of the early stage of the
chemical evolution of the Galaxy. Many high-resolution
spectroscopic abundance studies for metal-poor stars have been
carried out \citep[e.g.][]{mac95, rnb96, ful00, burris00,
gratton03, cayrel04,honda04, aoki05, francois07} and their results
have shown that the abundance patterns of metal-poor stars differ
from those of the Sun. In particular, it has been established that
the ratio of $\alpha$-elements to iron ratio in stars with
[Fe/H]$\lesssim-1.0$ is greater than the solar ratio by $\sim+0.4$
dex \citep{wheeler89}. This indicates the complex nature of the
chemical enrichment.

The general explanation for the chemical evolution of the Galaxy
has been based on the products of the two main types of supernovae
\citep{tinsley80}. The first, comprising the Type II SN and Type
Ib/Ic SNe which are the core-collapse-induced explosions of
massive stars. The second, the Type Ia SN is the thermonuclear
explosion of an accreting white dwarf in a close binary system.
The core-collapse SNe breakout shortly after early Galaxy
formation because their progenitors are massive stars with short
lifetimes, while Type Ia SNe occur after completing at least one
stellar lifetime of $\sim10^9$yr(for an 8$M_\odot$ star) resulting
in the formation of a white dwarf \citep{chiosi92}.
$\alpha$-elements are the main products of Type II and Type Ib/Ic
SNe while Fe is mainly produced by Type Ia SNe. Therefore the
ratio of $\alpha$-elements to Fe ratio is high in the halo stars
that were formed in the early stage of the Galaxy however this
ratio reduces to the solar value as observed in the trends of
$\alpha$-elements along metallicity.

However, the primary nucleosynthetic sources for most elements are
not well known. A relative abundance ratio, for instance, the
ratio of an element to Fe is a good tool for investigating the
product site of each element. However the inhomogeneous abundance
results of previous studies, which showed large scatters and/or
some shifts, rule out the use of this technique for this purpose.
In order to determine the abundance of a given element, the
atmospheric parameters (temperature \Teff, surface gravity $\log
g$, micro-turbulent velocity $\xi_v$, and overall metallicity
represented by [Fe/H]) need to be derived. Therefore differences
in these parameters result in differences in the results of
abundance analysis. The choice of the lines used for analysis and
the oscillator strength value ($\log gf$) adopted for deriving the
abundance also results in uncertainties and systematic differences
in the abundance results. A single-lin approximation is found to
be in adequate for some spectral lines of odd atomic number
elements with significant hyperfine structures \cite{rnb96}. It
has been shown that ignoring the hyperfine structure could lead to
abundance errors of up to 0.6 dex. Therefore, we aim to obtain
abundances of elements in stars with wide metallicity ranges as
accurately as possible by using homogeneous methods.

We divided the elements into three groups, namely, light elements
($Z\le20$), iron-peak elements ($21\le Z\le30$), and $n$-capture
elements ($Z\ge38$). Light elements are good tools for judging the
surface chemical contamination due to the internal mixing in
giants. Iron-peak elements which are mainly produced by explosive
Si-burning can give important information on the supernova
explosions. The iron-peak elements particularly in metal-poor
stars with [Fe/H]$\le-1.0$ are produced in massive core-collapse
supernovae \citep{tinsley80} and provide important constraints for
models of type II SNe.

The $n$-capture process is responsible for the production of heavy
elements beyond the iron-peak elements, however, its site has not
yet been determined with certainty. In particular, the
nucleosynthesis process of light $n$-capture elements is proposed
to be a combination of various processes \citep{montes07}.
Further, radioactive elements with sufficiently long half-lives,
such as uranium (U, Z=92) and thorium (Th, Z=90) can provide a
constraint on the age of metal-poor stars. In order to determine
the age of old stars with unstable radioactive elements, a stable
element synthesized in the same event is required to be set as
reference.

The transition probabilities of rare-earth $n$-capture elements
and their stable isotopes such as La, Sm, Eu, and Hf have recently
been measured with very high accuracy by Lawler and associates
\citep[e.g.][]{hartog, lawler01a, lawler01b, lawler07}. This has
enabled the derivation of more reliable $n$-capture element
abundance, which may provide new insights on the roles of the
$n$-capture process in the initial burst of Galactic
nucleosynthesis.

\section{Observations and reductions}

\subsection{Observations}
We selected the target stars for our study based on the following
criteria. Firstly, due to the small aperture (1.8 m) of Bohyunsan
Optical Astronomical Observatory(BOAO), we selected stars with a
visual magnitude brighter than 10 mag. Secondly, metal-poor stars
whose metallicity was determined as [Fe/H]$\le-1.0$ by previous
spectroscopic studies were selected. Thirdly, stars that are known
to have a high proper-motion but do not have previous abundance
studies were also included in our target set. We excluded K- and
M-type stars that have strong molecular lines as well as metallic
lines that are too strong and blended with other lines.

We have obtained high-resolution echelle spectra for 35 metal-poor
stars using the Bohyunsan Optical Echelle Spectrograph (BOES)
\citep{kim02} in two runs scheduled in November 2005 and April
2006.
BOES is a fiber-fed echelle spectrograph mounted on a 1.8-meter
telescope at BOAO (Bohyunsan Optical Astronomical Observatory).
The BOES detector is an EEV-CCD with $2048\times4102$ pixels and a
pixel size of $15\mu m$. The BOES covers a wide wavelength
range(3800 to 10500\AA) in a single exposure and has a high
efficiency($\sim15\%$). There are three kinds of fiber sets, with
apertures of 80$\mu m(1.1\arcsec)$, 200$\mu m(2.9\arcsec)$, and
300$\mu m(4.2\arcsec)$ corresponding to spectral resolving powers
of $R\simeq90000$, $44000$, and $30000$, respectively.

We observed most of the targets with the largest aperture(300$\mu
m$) in order to minimize light loss and obtain spectra with high
signal-to-noise ratio  because of the poor weather conditions at
BOAO(The seeing condition was $\sim3\arcsec$ in the first run
while that on the second run was $\sim4\arcsec$) during the
observing runs. Six stars(HD 94028, HD93487, HD 101227, HD 118659,
HD 148816, and HD159482) were observed using an aperture of
200$\mu m(2.9\arcsec)$, which gives a better resolution. We
observed no significant difference on comparing the spectra
observed from different apertures. A resolving power of $R\sim30k$
was found to be satisfactory for conducting abundance analysis. We
also observed B-type stars on every night to obtain telluric line
references.

We obtained bias frames for zero-correction and tungsten-halogen
lamp(THL) frames for flat-fielding, and Th-Ar spectra for
wavelength calibration. Since dark current of BOES CCD is very
small($\sim2e^- $/hr/pixel) relative to read-out
noise($\leq3e^-$), it is not necessary to obtain dark frames. The
observation log is presented in Table.\ref{table:log}

\subsection{Reductions}
The reductions were performed by following the standard steps for
echelle spectrum using IRAF. The biases were first corrected by
subtracting the median combined bias frame using the IRAF/ccdred
task. The dark currents were not corrected because the dark
current of the BOES CCD is negligibly small.

We adapted a technique for aperture extractions and flattening
that differs slightly from the standard reduction procedures. We
first prepared a master flat \emph{image} by combining
preprocessed flat images. This master flat image is used only as
aperture tracing reference. Using this master flat image, we
defined apertures and background regions to be used for scattered
light subtraction using the IRAF/ECHELLE/apall task. Subsequently,
flat spectra and object spectra were extracted. We then produced
the master flat\emph{spectrum} by combining the flat spectra using
the IRAF/ECHELLE/scombine task.

The flat-field correction was carried out by dividing each
extracted order of the object spectra by the corresponding order
of the master flat \emph{spectrum}. The blaze function is also
corrected using this procedure. Since the THL lamp is not a
perfect flat source, the flux of the flat-corrected spectra
becomes distorted, however this distortion can be corrected by
continuum normalization. Compared to the standard echelle
reduction procedure, this procedure is faster and simple and it
corrects the blaze function prior to the continuum normalization.
For the case wherein the information of the flux is not required
and only the continuum-normalized spectrum is needed--usually for
equivalent width(EW) measurements--this procedure would be more
efficient for the reduction process.

\begin{figure}[!ht]
\plotone{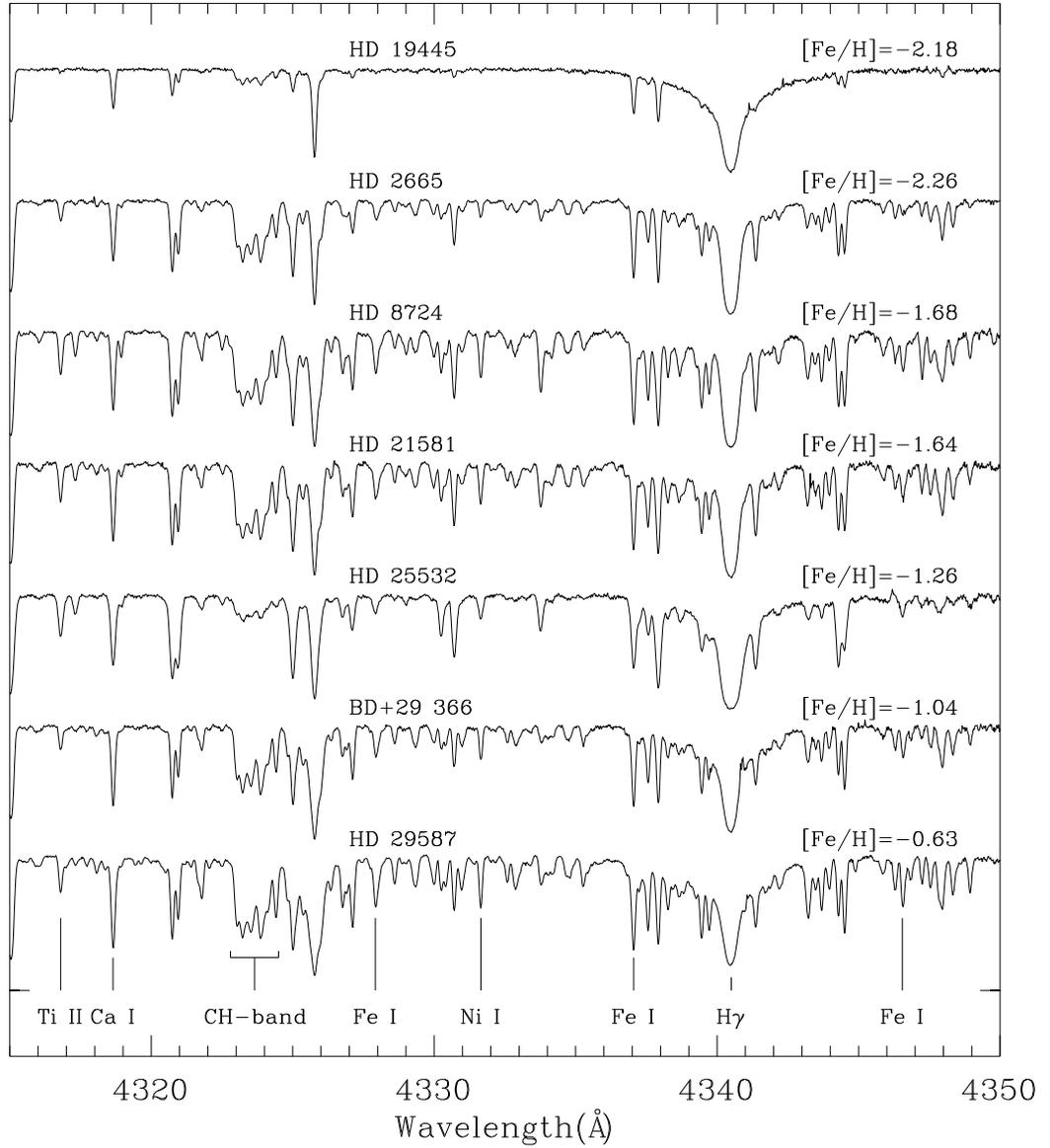} \caption{Examples of spectra for seven stars over
a wavelength range of 4315-4350\AA. Several identified lines are
indicated. This spectral region includes the CH molecular band
used for deriving carbon abundance. H$\gamma$ line is observed at
4340\AA.} \label{fig:sample}
\end{figure}

The wavelength calibration function was obtained using the
IRAF/ECHELL/ecidentify task based on the Th-Ar lamp spectrum
corresponding to the first night of the observing run. The daily
variation in the wavelength of BOES was very small($<0.001\AA$),
therefore, the IRAF/ECHELL/ecreidentify task was utilized for the
remaining nights of the run. Figure.\ref{fig:sample} shows a small
section of the sample spectra around the $H\gamma$ line, with a
wavelength range of 4320 to 4350 \AA~for seven stars whose
metallicities [Fe/H] range from $-0.63$ to $-2.26$ dex.

\subsection{Equivalent Widths}
Equivalent widths for most elements were measured using Gaussian
fitting in the IRAF/splot task. Local continuum was defined by
visual inspection for each line. Weak Fe I and Fe II lines with
log(EW/$\lambda$)$< -4.9$ or EWs less than 100m\AA~were used to
determine the atmospheric parameters and to carry out abundance
analysis, as strong lines lie in the over-saturation region of the
curve of growth, where Gaussian fitting is inadequate and an
accurate damping constant is required. However, when only strong
lines were available for abundance analysis, we measured EWs by
means of voigt profile fitting. The sources of adopted atomic
information are listed in Table.\ref{table:loggf}. A small sample
of measured equivalent widths, along with the corresponding atomic
information, is presented in Table.\ref{table:ew}.

\begin{figure}[!ht]
\plotone{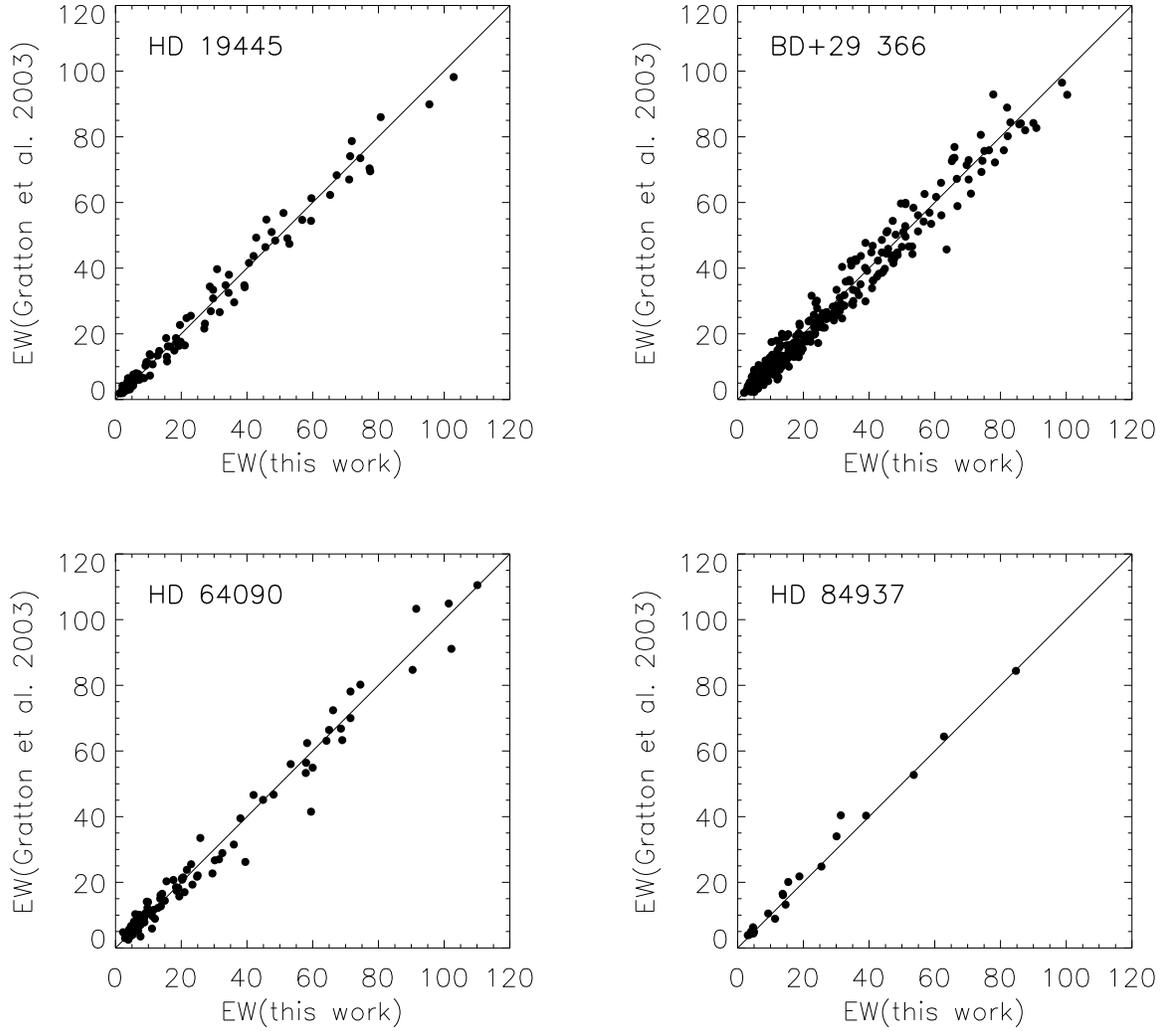} \caption{Comparison between the EWs(in $m\AA$)
measured from the BOES spectra and those corresponding to
\cite{gratton03}} \label{fig:ew_comp1}
\end{figure}

\begin{figure}[!ht]
\plotone{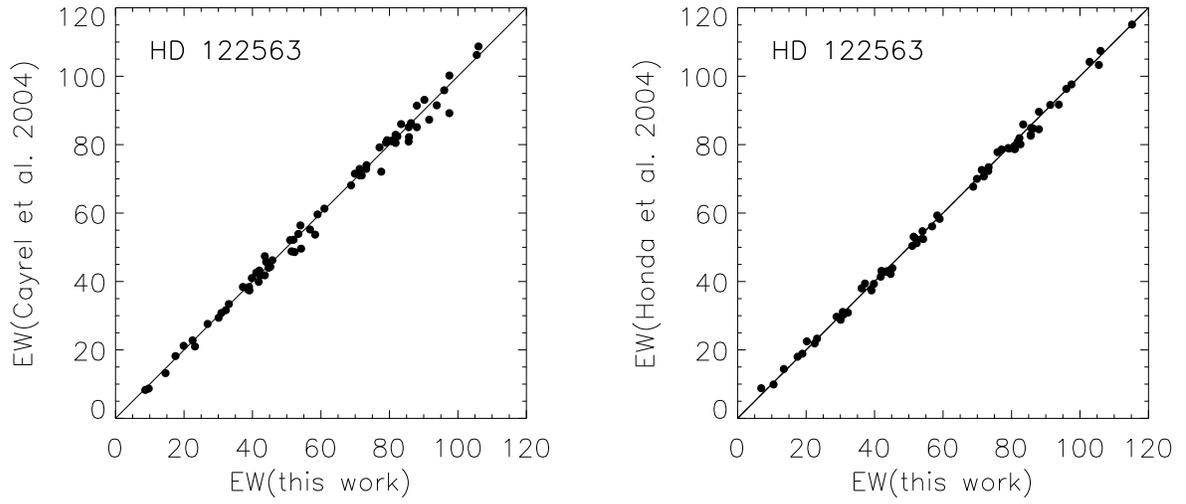} \caption{Comparison between the EWs measured from
the BOES spectra and those corresponding to \cite{cayrel04} and
\cite{honda04} for HD 122563} \label{fig:ew_comp2}
\end{figure}

We compared the measured EWs with those of previous studies. A
graphical comparison for the common stars is shown in Figure.
\ref{fig:ew_comp1} and Figure. \ref{fig:ew_comp2}. Four stars,
namely, BD+29 366, HD 19445, HD 64090, and HD 84937 are common
between our study and \cite{gratton03}. The mean differences (this
work minus \cite{gratton03}) are $0.50~\pm~0.23~$m\AA~(sdom
\footnote{sdom, standard deviation of mean
$=\sqrt{\frac{\sigma^2}{N}}$}, 320 lines) for BD+29 366,
$0.03~\pm~ 0.29~$m\AA(110 lines) for HD 19445,
$0.46~\pm~0.46~$m\AA~(96 lines) for HD 64090, and
$-1.36~\pm~0.57~$m\AA~ (20 lines) for HD 84937. Figure.
\ref{fig:ew_comp2} shows the comparison of EWs of HD 122563 with
those of \cite{cayrel04} and \cite{honda04}. The mean differences
corresponding to \cite{cayrel04} and \cite{honda04} are
$0.25~\pm~0.27~$m\AA~(78 lines) and $0.33~\pm~0.18~$m\AA~(67
lines), respectively. There is no significant offset observed in
the measurement of EWs.

The expected EW uncertainty derived from Cayrel's formula (1988)
for our spectra with $S/N\sim200$ is $\sim2$m\AA. This uncertainty
ignores the measurement errors that occur in continuum definition
and result from the departure of the Gaussian profile of the
observed line. For weak lines, the departure from Gaussian line
profile is negligible, while the major error in EW measurements
results from the local continuum definition. We repeated the EW
measurements for some well isolated lines by setting different
local continuum to obtain an error of less than 4m\AA, however, we
found that the error in EW depends on the line strength and
signal-to-noise ratio of a spectrum.

\subsection{Radial Velocities}

\begin{figure}[!ht]
\plotone{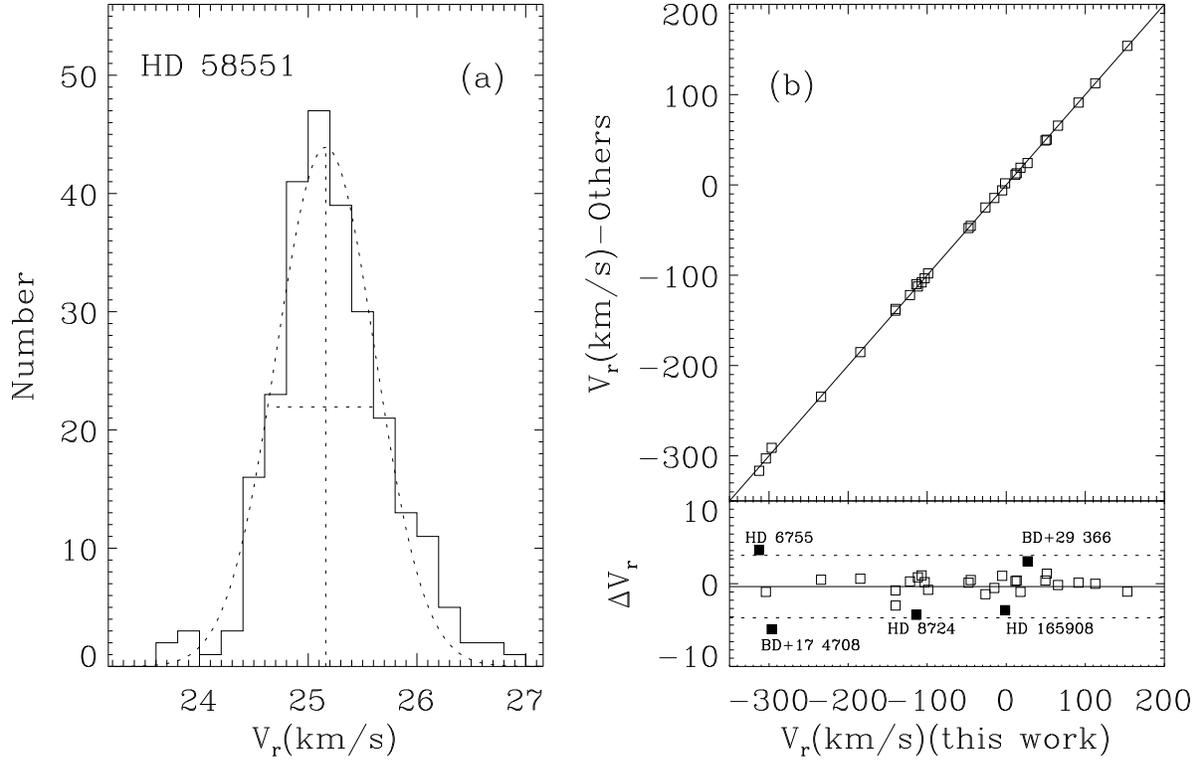} \caption{(a) An example of the determination of
radial velocity of HD 58551 prior to heliocentric velocity
correction. The dotted line is the Gaussian fitting result.
(b){\itshape(Upper panel)} Comparison between radial velocities
measured from the BOES spectra and those from other radial
velocity studies. {\itshape(Lower panel)} Differences in radial
velocities. The solid line represents the mean offset and the
dotted line represents a $2\sigma$ dispersion of velocity
differences. The velocities of HD 6755, HD 8724,  HD 165908, and
BD+17 4708, represented as filled squares, show large deviations
from previous studies. These stars, except HD 8724, are known as
binaries.} \label{fig:rad_vel}.
\end{figure}

Radial velocities were measured from the observed centers of the
Fe I and Fe II lines that are well-isolated and have well-defined
Gaussian profiles. The distribution of radial velocities derived
from each line for a star was Gaussian-fitted to obtain the
stellar velocity, as Figure. \ref{fig:rad_vel}(a). The radial
velocities for program stars are presented in the last column of
Table.~\ref{table:atpar}. The typical error of radial velocitiy
($2\sigma$ dispersion in the Gaussian profile) is $\sim0.45$ km/s.
A comparison of the radial velocities obtained in this work and
those of previous studies \citep{geneva04, latham02} is presented
in Figure. \ref{fig:rad_vel}(b) in order to demonstrate the a good
agreement between the two.

The mean offset is $+0.51~\pm~0.66$ km/s (sdom, N=31), indicted as
a solid line in the lower panel of Figure.\ref{fig:rad_vel}(b),
the dotted line indicates a $2\sigma$ dispersion. Among the 35
stars under consideration, two stars, namely, HD 6755 and BD+17
4708, exhibit velocity differences greater than the 2$\sigma$
dispersion, while HD 8724, HD 165908, and BD+29 366 exhibit the
velocity differences close to the $2\sigma$ dispersion. These are
indicated as filled squares in the lower panel of Figure
\ref{fig:rad_vel}(b). BD+17 4708 and HD 6755 are known as binaries
in previous radial velocity studies of \cite{latham02} and
\cite{carney03}, and our radial velocities are in good agreement
with the radial velocity curves calculated from their orbital
solutions. HD 165908 is also a star in a binary system(b Her).
BD+29 366 is a suspected binary with an unknown period
\citep{latham02}. Only one previous radial velocity study
\citep{barbier89} exists in relation to HD 8724. Considering its
error of 1 km/s and a velocity difference of approximately 2.5
km/s from our results, we suspect that HD 8724 may be a binary
system, which requires further radial-velocity study.

\section{Atmospheric parameters}

In order to derive elemental abundances, an appropriate model
atmosphere must be constructed prior to conducting abundance
analysis. We used Kurucz model atmosphere grids computed with the
new opacity distribution function(ODF)(http://kurucz.harvard.edu).
A Linux ported ATLAS program \citep{sbordone} was used to
interpolate model atmospheres for the atmospheric model between
grid points. For the metal-poor stars with [Fe/H]$<-0.5$,
$\alpha$-element enhanced models ([$\alpha$/Fe]=+0.4) were used.
The adopted $\log gf$ values for the iron lines are presented in
Table.~\ref{table:ew}. For the determination of atmospheric
parameters, we used only the Fe I and Fe II lines weaker than
log(EW/$\lambda$) $<-4.90$ or EW $< 100$m\AA~in order to avoid the
saturated lines. The LTE abundance analysis program, MOOG,
\citep{sneden73} was used to derive Fe I and Fe II abundances.

\begin{figure}[!ht]
\plotone{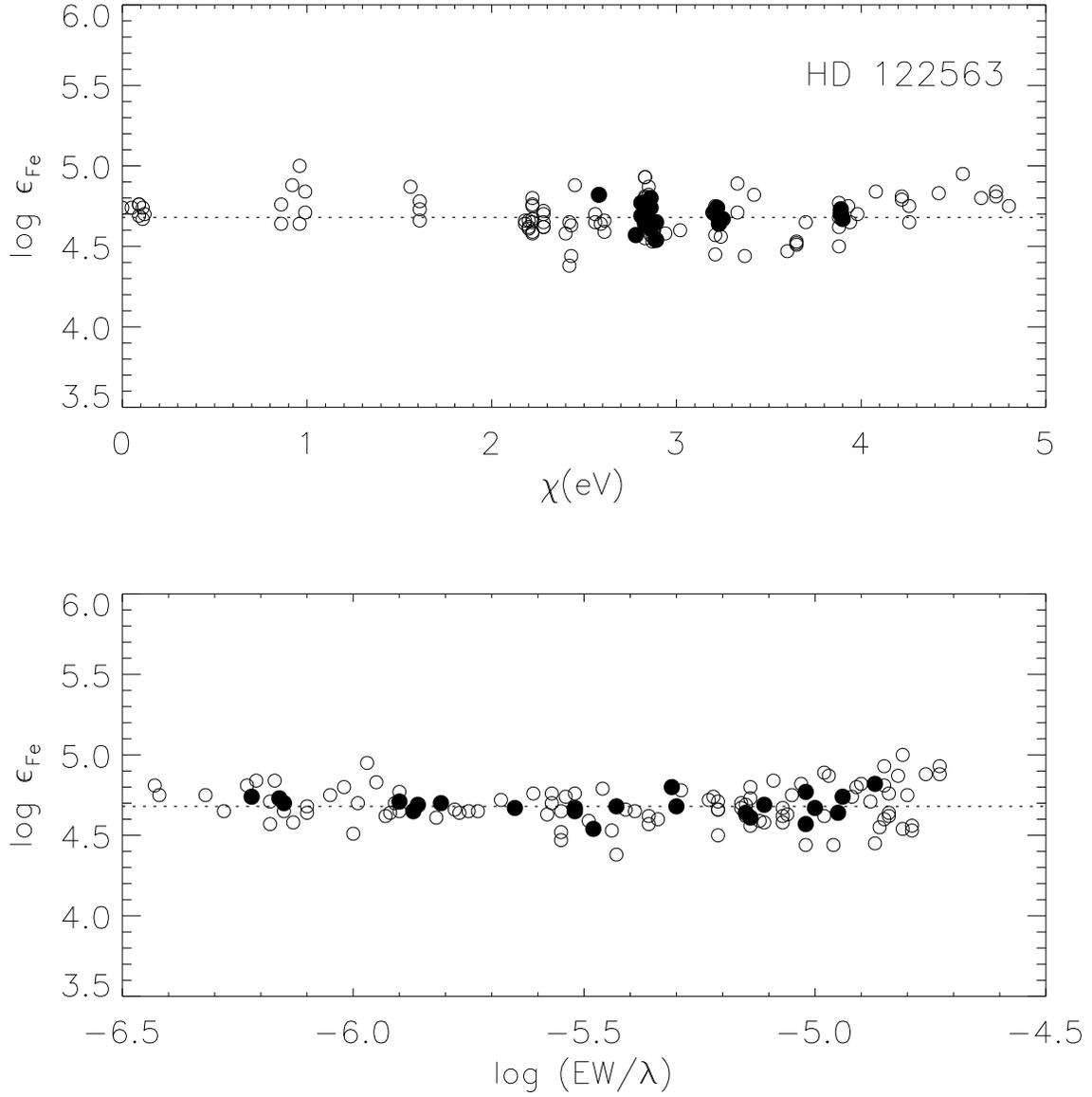} \caption{Fe abundances for HD 122563 with derived
atmospheric parameters(\Teff$=4430K$,  $\log g=0.58$, $\xi_v =
2.15$ km/s). Open circle shows iron abundance derived from the Fe
I line and filled circle shows that derived from the Fe II line.
}\label{fig:fe_abun}
\end{figure}

We determined the atmospheric parameters by following a
self-consistent LTE analysis technique. First, the value of
\Teff~was adjusted so that there is no trend between the iron
abundances from each of the Fe I lines and their lower level
excitation potentials. The micro-turbulent velocity($\xi_v$) was
then determined to exhibit no correlation with the Fe I line
strength and the iron abundance derived from each lines. Finally,
the value of log$g$~ was adjusted so that the iron abundance
derived from the Fe I lines matches the iron abundance given from
the Fe II lines. This process was iterated until the determined
parameters (\Teff, log$g$, and $\xi_v$) satisfied the above
criteria simultaneously. We used more than 100 Fe I lines and a
few tens of Fe II lines to determine the atmospheric parameters.
Figure. \ref{fig:fe_abun} shows the Fe abundance of HD 122563 with
the optimal atmospheric parameters to be set as \Teff$= 4430$K,
$\log g=0.58$, $\xi_v=2.15$km/s, and $\log\epsilon_{Fe}=4.68$. The
derived atmospheric parameters derived for our program stars are
presented in Table \ref{table:atpar}.

\begin{figure}[!ht]
\plotone{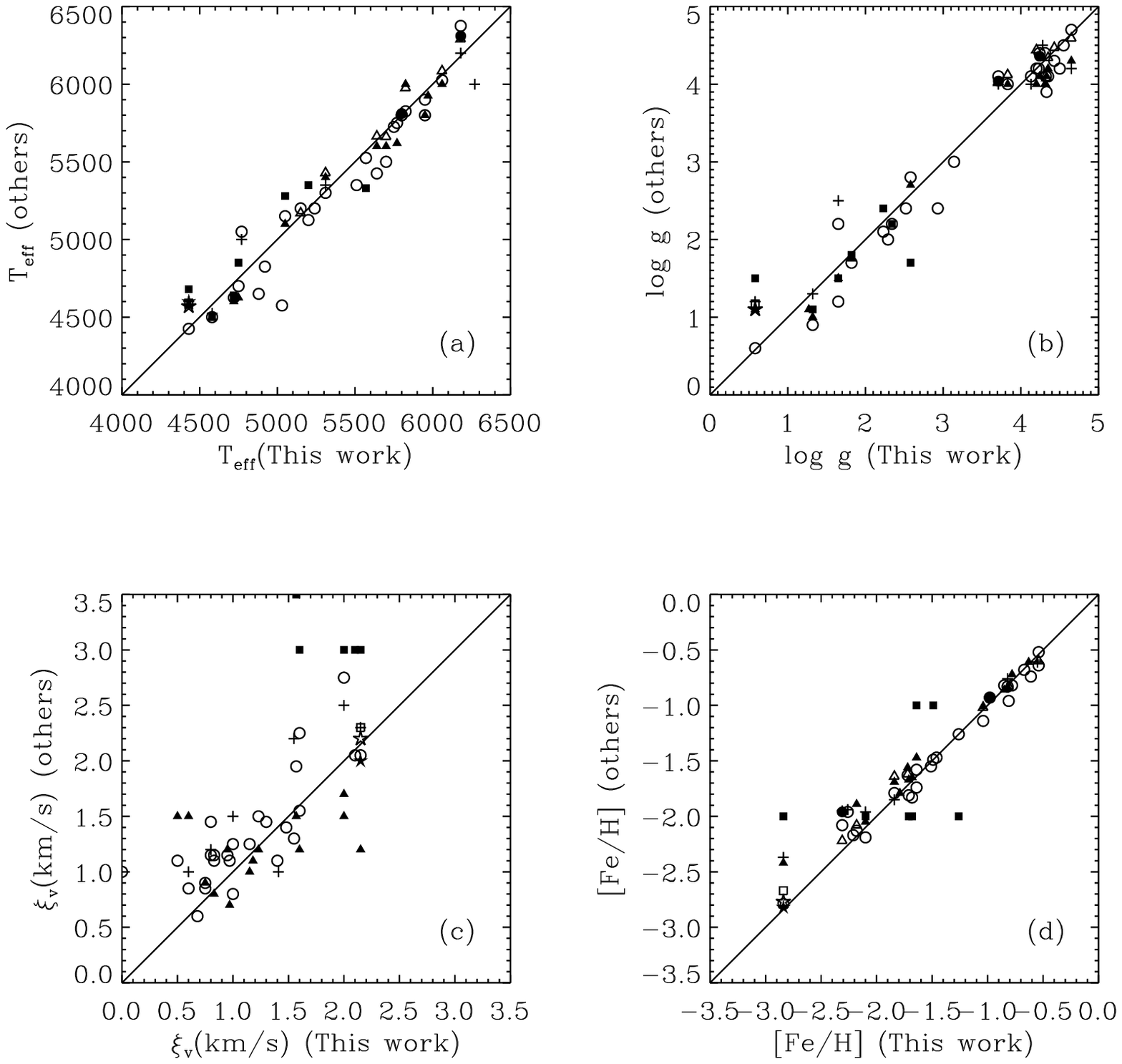} \caption{Atmospheric parameter(\Teff, $\log g$,
$\xi_v$, $[Fe/H]$) comparisons between this study and various
other studies.  Filled square indicates comparison with
\cite{go84}; plus, indicates comparison with \cite{sgc91}; open
square, \cite{gs94}; open circle, \cite{ful00}; filled triangle,
\cite{mk01}; open triangle, \cite{gratton03}; open star,
\cite{honda04}; filled star, \cite{cayrel04}; and filled circle,
\cite{johnsell05} }\label{fig:atcomp}
\end{figure}

We compared the atmospheric parameters derived in our study with
those obtained in other studies for common objects. This is
depicted in Figure \ref{fig:atcomp}, where each symbol represents
a comparison of our atmospheric parameters with those obtained in
other literature, using the following key. A filled square
represents comparison with \cite{go84}, a plus symbol indicates
comparison with \cite{sgc91}, open square, \cite{gs94}, open
circle, \cite{ful00}, filled triangle, \cite{mk01}, open triangle,
\cite{gratton03}, open star, \cite{honda04}, filled star,
\cite{cayrel04}, and filled circle, \cite{johnsell05}. We have
examined whether a systematic difference exists in the atmospheric
parameters depending on the method employed. The atmospheric
parameters in \cite{ful00} were derived in essentially the same
manner as in our study, with the exception that the metal
abundance of the model atmosphere was enhanced in \cite{ful00} in
order to compensate for the increased electron density caused by
$\alpha$-element enhancement in metal-poor stars.

As shown in Figure. \ref{fig:atcomp} (a), the effective
temperatures obtained in this work are in good agreement with
those of other studies. The \dTeff~ or the mean of
$\Delta$\Teff=\Teff~(this study)$-$\Teff~(other studies) between
our study and \cite{ful00}, wherein \Teff~was determined using the
same method, is $+54\pm26K$ (sdom, N=26). The value of \dTeff~for
\cite{mk01}, wherein \Teff~was derived using wing profile fitting
of H$\alpha$, is $+37\pm29K$ (sdom, N=14) and that for
\cite{gratton03}, wherein \Teff~was obtained by employing a
photometric method using (\bv) and $(b-y)$ color index, is $-59\pm
26K$ (sdom, N=7). There are no significant differences in
\Teff~based on the different methods employed. In general, errors
of up to $100K$ in effective temperature are acceptable.

Figure. \ref{fig:atcomp} (b) shows the comparison of the log $g$
values obtained in this study with those obtained in other
studies. The mean offset in log $g$ is $+0.08\pm0.05$ in the case
of \cite{ful00} and $+0.11\pm0.06$ in \cite{mk01}, wherein surface
gravity was determined using the ionization balance between Fe I
and Fe II lines. In \cite{gratton03}, surface gravities were
obtained using absolute visual magnitudes deduced from Hipparcos
parallexes and masses obtained by interpolating the position of
the star along the isochrones. The typical values of their errors
in surface gravity, resulting mainly from uncertainties in the
parallexes, is approximately $\pm0.1$ dex. The mean difference in
surface gravities in \cite{gratton03} is $-0.08\pm0.08$ dex. Thus
no significant differences are observed. The non-LTE effect may
result in a difference between the surface gravities determined by
the photometric method and the spectroscopic method, which is
based on the LTE assumption involving ionization balance. This
non-LTE effects are expected to be significant under the low
surface gravity condition. Comparison of common objects common
between this study and \cite{gratton03}, whose program stars are
subdwarfs or early subgiants, shows no significant difference.

Figure. \ref{fig:atcomp} (d) present a comparison of the iron
abundance obtained in this study with that obtained in other
studies. While \cite{ful00} used the same method for deriving
atmospheric parameters, the value of [Fe/H] in the atmospheric
model used in \cite{ful00} was set marginally higher than the
derived iron abundance in order to compensate for the electron
density due to $\alpha$-elements enhancement in metal-poor stars.
Using an $\alpha$-enhanced ($+0.4$dex) model in our study, we
compared the [Fe/H] value of our model to that of \cite{ful00} and
observed no difference ($0.00\pm0.02$~dex). There are no
significant systematic errors in [Fe/H], however, our results are
metal deficient by $0.06\pm0.03$~dex as compared to
\cite{gratton03}, and by $0.10\pm0.03$~dex as compared to
\cite{mk01}.

\section{Summary}
We have obtained high-resolution ($R\simeq30$k), high
signal-to-noise($S/N\geq100$) spectra of 35 metal-poor stars using
the BOES at BOAO, with a wavelength range of 3,800 to 10,500\AA.

EWs were measured using the Gaussian fitting method, and the local
continuum for each line was determined by visual inspection. In
cases where only strong lines were available for a given element,
EWs were measured by the voigt profile fitting method. Thus, we
presented the EWs of various elements. The measured EWs are in
strong agreement with those of previous studies.

Radial velocities were determined with an error of $\sim 0.45$
km/s from the Fe lines that were isolated and well-defined as
Gaussian profiles. HD 8724 is suspected to be a binary system. The
atmospheric parameters were determined by following a
self-consistent LTE analysis technique.

\Teff~was chosen such that Fe I lines of different excitation
potentials yield the same abundance and surface gravity was
determined from the ionization balance of Fe I and Fe II. The
micro-turbulent velocity was set to exhibit no correlation between
Fe abundance and the strength of the Fe I line. Model atmospheres
were constructed using Kurucz LTE atmospheric model calculated
with the new ODF. The atmospheric parameters we obtained are in
good agreement with those of previous studies despite the other
studies adapting methods different from ours to obtain these
parameters.

\clearpage

\begin{deluxetable}{rlccccrr}
\tablecolumns{8}\tabletypesize{\footnotesize} \tablewidth{0pt}
\tablecaption{Observational Log} \tablehead{ \colhead{Num.} &
\colhead{Object} & \colhead{R.A.(J2000)} & \colhead{Dec.(J2000)} &
\colhead{V} & \colhead{Observation date}& \colhead{$t_{exp}$}&
\colhead{$S/N$} } \startdata 1 & HD 2665 &   $ 00 ~~  30  ~~ 45.44
$ &   $   +   57 ~~ 03  ~~  53.6    $   & 7.65 & Nov. 2005 & 1800
& 250 \nl

2 & HD 6755 &   $   01 ~~  09  ~~ 43.00   $   &   $ + 61 ~~  32 ~~
30.2    $   & 7.68    & Nov. 2005   &   1800 & 250 \nl

3 & HD 8724 & $ 01 ~~ 26  ~~ 17.59   $   & $ +   17 ~~ 07 ~~  35.1
$ & 8.30 & Nov. 2005   &   2000    & 250 \nl

4 & BD+29 0366  &   $ 02 ~~ 10 ~~ 24.50 $ &   $   +   29 ~~ 48 ~~
23.7 $ & 8.79 & Nov. 2005 & 2400 & 240 \nl

5 & HD 19445 & $ 03 ~~ 08 ~~ 25.58   $   & $ + 26  ~~ 19 ~~ 51.4 $
& 8.04 & Nov. 2005 & 2400 & 260 \nl

6 & HD 21581 &   $ 03 ~~  28 ~~ 54.49 $ & $   - 00  ~~ 25  ~~ 03.1
$ & 8.70 & Nov. 2005 &   2400 & 220 \nl

7 & HD 25532 & $ 04 ~~ 04 ~~ 11.01 $ &   $   + 23 ~~ 24 ~~ 27.1 $
& 8.18 & Nov. 2005   & 2400 & 220 \nl

8 & HD 29587 & $ 04 ~~ 41 ~~ 36.32 $ & $ +   42 ~~ 07  ~~ 06.5 $ &
7.28 & Nov. 2005   & 2400 & 300 \nl

9 & BD+37 1458 & $ 06 ~~ 16 ~~ 01.52 $   & $ + 37  ~~ 43 ~~ 18.7 $
& 8.92 & Nov. 2005 & 2400    & 160 \nl

10 & HD 45391 & $ 06 ~~ 28 ~~ 46.03   $   & $ + 36 ~~  28  ~~ 47.9
$ & 7.11 & Nov. 2005 & 2400 & 280 \nl

11 & HD 58551 &   $ 07 ~~ 26 ~~ 50.25 $   & $ + 21 ~~ 32 ~~ 08.3 $
& 6.53 & Nov. 2005 & 600 & 300 \nl

12 & HD 59374 & $ 07 ~~ 30 ~~ 29.02 $ &   $ + 18 ~~ 57 ~~ 40.6 $ &
8.51 & Nov. 2005 & 3000 & 230 \nl

13 & HD 64090    & $ 07  ~~ 53 ~~ 33.12 $ & $ + 30 ~~ 36 ~~ 18.2 $
& 8.27 & Nov. 2005 & 3600 & 250 \nl

14 & HD 63791 & $ 07 ~~  54 ~~ 28.70 $ & $ + 62  ~~ 08 ~~ 10.8 $ &
8.20 & Nov. 2005   & 3600 & 250 \nl

15 & HD 73394 & $ 08  ~~ 40 ~~ 22.54 $ & $ + 51 ~~ 45  ~~ 06.6 $ &
7.71 & Apr. 2006 & 2400 & 150 \nl

16 & HD 84937 & $ 09 ~~ 48 ~~ 56.00 $ & $ + 13  ~~ 44 ~~ 39.3 $ &
8.29 & Nov. 2005 & 1800 & 250 \nl

17 & HD 237846 &   $ 09 ~~ 52 ~~ 38.68 $ & $ + 57 ~~ 54 ~~ 58.6 $
& 9.94 & Nov. 2005 & 7200 & 200 \nl

18 & HD 93487 & $ 10 ~~ 47 ~~ 55.53 $ & $ + 23 ~~ 20 ~~ 07.0 $ &
7.71 & Apr. 2006 & 4800 & 130 \nl

19 & HD 94028 & $ 10 ~~ 51 ~~ 28.12 $ & $ + 20 ~~ 16 ~~ 39.0 $ &
8.23 & Apr. 2006 & 3000 & 150 \nl

20 & HD 101227 & $ 11 ~~ 39 ~~ 06.22 $ & $ + 44 ~~ 18 ~~ 20.3 $ &
8.39 & Apr. 2006 & 4800 & 175 \nl

21 & HD 105546 & $ 12 ~~ 09 ~~ 02.72 $ & $ + 59 ~~ 01 ~~ 05.1 $ &
8.61 & Apr. 2006 & 3600 & 130 \nl

22 & HD 114095 & $ 13 ~~ 08 ~~ 25.79 $   & $ - 07 ~~ 18 ~~ 30.5 $
& 8.35 & Apr. 2006 & 4000 & 120 \nl

23 & HD 114762 & $ 13 ~~ 12 ~~ 19.74 $   & $ + 17 ~~ 31 ~~ 01.6 $
& 7.30 & Apr. 2006 & 2400 & 170 \nl

24 & HD 118659 & $ 13 ~~ 38 ~~ 00.47 $ & $ + 19 ~~ 08 ~~ 53.1 $ &
8.84 & Apr. 2006 & 4800 & 110 \nl

25 & HD 122563 & $ 14  ~~ 02 ~~ 31.85 $ & $ + 09 ~~ 41 ~~ 09.9 $ &
6.20 & Apr. 2006 & 1800 & 220 \nl

26 & HD 123710 & $ 14 ~~ 04 ~~ 57.07 $   & $ + 74 ~~ 34 ~~ 24.9 $
& 8.21 & Apr. 2006 & 3600 & 160 \nl

27 & HD 148816 & $ 16 ~~ 30 ~~ 28.46 $   & $ + 04 ~~  10 ~~ 41.0 $
& 7.27 & Apr. 2006 & 3000 & 110 \nl

28 & HD 159482 & $ 17 ~~ 34 ~~ 43.06 $ & $ + 06 ~~ 00 ~~ 51.6 $ &
8.39 & Apr. 2006 & 4800 & 160 \nl

29 & HD 165908 & $ 18 ~~  07 ~~ 01.54 $ & $ + 30 ~~ 33 ~~ 43.7 $ &
5.07 & Apr. 2006 & 300 & 200 \nl

30 & HD 175305 & $ 18 ~~ 47 ~~ 06.44 $ & $ + 74 ~~ 43 ~~ 31.4 $ &
7.20 & Apr. 2006 & 2400 & 260 \nl

31 & HD 201889 & $ 21 ~~ 11 ~~ 59.52 $ & $ + 24 ~~ 10 ~~ 05.0 $ &
8.04 & Nov. 2005 & 1800 & 240 \nl

32 & HD 204543 & $ 21 ~~  29 ~~ 28.21 $ & $ - 03 ~~ 30 ~~ 55.4 $ &
8.31 & Nov. 2005 & 2400 & 290 \nl

33 & BD+22 4454 & $ 21 ~~ 39 ~~ 36.46 $ & $   + 23 ~~ 15 ~~ 55.9 $
& 9.50 & Nov. 2005 & 1800 & 140 \nl

34 & BD+17 4708 & $ 22 ~~ 11 ~~ 31.37 $ & $ + 18 ~~ 05 ~~ 34.2 $ &
9.47 & Nov. 2005 & 5400    & 120 \nl

35 & HD 221170 & $ 23 ~~ 29 ~~ 28.81 $ & $ + 30 ~~ 25 ~~ 57.8 $ &
7.71 & Nov. 2005 & 1200 & 260 
\enddata
\label{table:log}
\end{deluxetable}

\begin{deluxetable}{clrrrrrrrrrrrrrr}

\tabletypesize{\scriptsize} \tablecolumns{15} \tablewidth{0pt}
\tablecaption{Equivalent Widths} 
\tablehead{
\colhead{} & \colhead{}& \colhead{}& \colhead{}
&  \multicolumn{11}{c}{Equivalent Width (m\AA)} \\
\cline{5-15} \\
\colhead{Wavelength} &
\colhead{Species} & \colhead{eV} &\colhead{$\log gf$} &\colhead{1}
&\colhead{2} &\colhead{3} &\colhead{4} &\colhead{5} &\colhead{6}
&\colhead{7} &\colhead{8} &\colhead{9} &\colhead{10} &\colhead{11}
}
\startdata
6335.34     &   Fe I    &   2.20    &   -2.27   &   51.2    &
67.4    &   90.9    &   61.9    &   10.2    &   73.7    &   61.0
&   78.3    &   25.7    &   80.6    &   62.9    \nl 6380.75     &
Fe I    &   4.19    &   -1.37   &   3.7     &   9.7     &   16.8
&   15.8    &       &   12.6    &   11.5    &   30.1    &       &
33.3    &   20.7    \nl 6392.54     &   Fe I    &   2.28    &
-3.97   &       &   3.8     &   7.4 &   3.8 &       &   3.5 &
&   6.6     &       &   7.4     &       \nl 6393.61     &   Fe I
&   2.43    &   -1.43   &   73.6    &   85.9    &   108.4   &
85.3    &   23.4    &   92.6    &   88.4    &   106.4   &   45.6
&   111.9   &   85.4    \nl 6411.66     &   Fe I    &   3.65    &
-0.60   &   41.2    &   61.7    &   78.8    &   74.5    &   12  &
68.1    &   65.8    &   102.4   &   27.5    &   109.8   &   79.0
\nl 6421.36     &   Fe I    &   2.28    &   -2.03   &   58.2    &
74.0    &   98.8    &   70.3    &   13.7    &   83.5    &   76.5
&   90.5    &   33.1    &   93.4    &   73.9    \nl 6481.88     &
Fe I    &   2.28    &   -2.98   &   15.8    &   33.4    &   50  &
23.3    &       &   32.9    &   28.6    &   50.8    &   8.9 &
47.0    &   24.0    \nl 6498.94     &   Fe I    &   0.96    &
-4.70   &   38.5    &   17.8    &   42.2    &   13.5    &   13.8
&   24.6    &   11.1    &   26.8    &       &   28.3    &   9.2
\nl 6518.37     &   Fe I    &   2.83    &   -2.46   &   7.2     &
18.5    &   30.9    &   24.5    &   1.8 &   23.8    &   19.5    &
38.7    &   3.4 &   39.3    &   20.5    \nl 6533.94     &   Fe I
&   4.56    &   -1.29   &   3.9     &       &   8.2 &   6.6 &
&   6.3 &   6.9     &   18.2    &       &   20.0    &   14.5
\nl 6574.25     &   Fe I    &   0.99    &   -5.00   &   6.2     &
&   36.8    &       &       &   14.6    &       &   15.8    &
&   16.1    &       \nl 6581.22     &   Fe I    &   1.49    &
-4.68   &       &       &   14  &   5   &       &   7.4 &   4.2
&   10.2    &       &   11.0    &   4.2     \nl 6593.88     &   Fe
I    &   2.43    &   -2.42   &   31.9    &   43.9    &   68.8    &
44.3    &       &   53.4    &   37.6    &   63.1    &   11.5    &
66.9    &   45.2    \nl 6608.04     &   Fe I    &   2.28    &
-3.96   &       &   3.1     &       &   3.1 &       &   4.6 &
&   7.1     &       &   8.7     &   3.4     \nl 6609.12     &   Fe
I    &   2.56    &   -2.69   &   12.6    &   24.9    &   43.4    &
24.7    &       &   32.2    &   18.4    &   43.9    &       &
47.2    &   26.1    \nl 6625.04     &   Fe I    &   1.01    &
-5.37   &       &   3.4     &   10.5    &   5.1 &       &   7.7 &
&   7.9     &       &   9.5     &       \nl 6627.56     &   Fe I
&   4.55    &   -1.50   &       &   4.0     &   2.9 &   4   &
&   3.4 &       &   12.0    &       &   12.4    &   8.5     \nl
6633.76     &   Fe I    &   4.56    &   -0.82   &   6.9     &
11.7    &       &   23.6    &   2.3 &   15.5    &   15.5    &
41.1    &       &   44.8    &   29.6    \nl 6703.58     &   Fe I
&   2.76    &   -3.01   &   3.9     &   8.5     &   14.9    &
7.9 &       &   10.3    &   5.7     &   18.8    &       &   19.5
&   8.4     \nl 6713.75     &   Fe I    &   4.80    &   -1.41   &
&       &   2.5 &   2.2 &       &   4.4 &       &   9.2     &
&   9.4     &   5.4   

\enddata
\tablecomments{Table \ref{table:ew} is presented in its entirety
in the electronic edition of the {\it Astrophysical Journal}. A
portion is shown here for guidance regarding its form and
content.}
\label{table:ew}
\end{deluxetable}

\begin{deluxetable}{llll}
\tablecolumns{2}\tabletypesize{\footnotesize} \tablewidth{0pt}
\tablecaption{References of adopted $\log gf$ value} \tablehead{ \colhead{Species} & \colhead{Reference} & \colhead{Species} &
\colhead{Reference} }

\startdata
Fe I    & Oxford group          & V I   & VALD\citep{kupka99}\\
        & \cite{bard91}         & V II  & VALD\citep{kupka99} \\
        & \cite{bard94}         & Cr I  & Oxford group \\ 
        & \cite{obrian91b}      & Mn I  & \cite{booth84}\\
Fe II   & \cite{holweger90}  	& Co I  & \cite{cardon82}  \\
        & \cite{biemont91b}		& Ni I & VALD\citep{kupka99} \\
        & \cite{hannaford92} 	& Cu I & \cite{bielsky75} \\
        & \cite{blackwell80}  	& Zn I & VALD\citep{kupka99} \\

Li I    & \cite{weiss63}        & Sr II & \cite{gs94}  \\
O I 	& \cite{biemont91}  		& Y II & \cite{hannaford82}  \\
$\rm{[O I]}$ & \cite{kurucz95}     & Zr II & \cite{biemont81} \\
Na I & \cite{fischer02}            & Ba II & \cite{Mcwilliam98}\\
Mg I & \cite{gratton03}      & La II & \cite{lawler01a}\\
Al I & \cite{mendoza95}      & Ce II & \cite{palmeri00} \\
Si I & \cite{obrian91}	  	& Nd II & \cite{hartog03} \\
K I & NIST                      & Sm II & \cite{lawler06} \\
Ca I & \cite{smith81}   & Eu II & \cite{lawler01b}\\
Sc II & NIST                    & Dy II & \cite{kusz92} \\
Ti I & Oxford group             & Hf II & \cite{lawler07}\\
Ti II & \cite{bizzarri93} 

\enddata
\label{table:loggf}
\end{deluxetable}

\begin{deluxetable}{rlrrrrrr}
\tabletypesize{\footnotesize} \tablecolumns{8}\footnotesize
\tablewidth{0pt} \tablecaption{Atmospheric parameters and radial
velocities of stars} \tablehead{ \colhead{Num} & \colhead{ID} &
\colhead{\Teff} & \colhead{$\log g$} & \colhead{$\xi_{v}$} &
\colhead{$\log \varepsilon_{Fe}$} & \colhead{[Fe/H]}&
\colhead{$V_{rad}$} }

\startdata

1 & HD 2665 & 4770    &   1.65    & 1.55    & 5.26    &$ -2.26 $&
$ -382.68 \pm 0.44 $ \nl

2 & HD 6755 &   5050    & 2.58 & 1.23    & 5.88 &$  -1.64   $&  $
-312.55     \pm 0.42 $ \nl

3 & HD 8724 &   4720 & 1.65    & 1.57 & 5.84    &$ -1.68 $&  $
-113.76 \pm 0.45 $ \nl

4 & BD+29 0366 & 5640 & 4.35 &   0.75 & 6.48 &$ -1.04   $& $ 26.91
\pm 0.33 $ \nl

5 & HD 19445    & 5825 & 4.20 & 0.50 & 5.34 &$ -2.18 $& $ -140.11
\pm 0.50 $ \nl

6 & HD 21581 & 4920 & 2.29 & 1.30 & 5.88 &$ -1.64 $&  $ 153.04 \pm
0.37    $ \nl

7 & HD 25532 &   5570 & 2.34 & 2.10    & 6.26 &$  -1.26 $&  $
-111.73 \pm 0.48 $ \nl

8 & HD 29587 & 5670 & 4.43 & 1.08 & 6.89 &$ -0.63   $& $ 112.59
\pm 0.38 $ \nl

9 & BD+37 1458  & 5240 & 3.14 & 1.00 & 5.31 &$  -2.21 $&  $ 242.60
\pm 0.38 $ \nl

10 & HD 45391 & 5720 & 4.50    & 1.05 & 7.00 &$ -0.52 $& $ -5.15
\pm 0.43 $ \nl

11 & HD 58551 & 6270 & 4.28 & 1.41 & 6.97 &$ -0.55 $& $ 51.21 \pm
0.45 $   \nl

12 & HD 59374    & 5800 & 4.24 & 0.83 & 6.54 &$ -0.98 $& $ 91.32
\pm 0.40 $ \nl

13 & HD 64090 & 5310 & 4.65    &   0.60 & 5.68 &$ -1.84 $& $
-234.22 \pm 0.48 $ \nl

14 & HD 63791    & 4750 & 1.82 & 1.60 & 5.81 &$ -1.71 $& $ -107.00
\pm 0.41    $ \nl

15 & HD 73394 & 4530 & 1.48 & 1.93 & 5.96 &$  -1.56 $& $ -101.23
\pm 0.39    $ \nl

16 & HD 84937 & 6180 & 3.71 & 1.00 & 5.21 &$ -2.31 $& $ -15.02 \pm
0.51 $ \nl

17 & HD 237846 & 4680 & 1.28 & 1.63 & 4.40 &$ -3.12   $& $ -303.99
\pm 0.39 $ \nl

18 & HD 93487 & 5350 & 2.45 & 1.90 & 6.47 &$ -1.05 $& $ 79.11 \pm
0.43 $ \nl

19 & HD 94028 & 5950 & 4.23 & 0.80 & 6.01 &$ -1.51 $& $ 65.44 \pm
0.41 $ \nl

20 & HD 101227 & 5550 & 4.51 & 0.90 & 7.10 &$ -0.42 $& $ 11.61 \pm
0.34 $ \nl

21 & HD 105546 & 5200 & 2.23 & 1.60 & 6.03 &$ -1.49 $& $ 18.01 \pm
0.38 $ \nl

22 & HD 114095 & 4880 & 2.93 & 1.40 & 6.91 &$ -0.61 $& $ 75.82 \pm
0.38 $ \nl

23 & HD 114762 & 5950 & 4.32 & 1.15 & 6.74 &$ -0.78   $& $ 49.58
\pm 0.39 $ \nl

24 & HD 118659 & 5510 & 4.50 & 0.68 & 6.85 &$ -0.67 $& $ -44.96
\pm 0.35 $ \nl

25 & HD 122563 & 4430 & 0.58 & 2.15 & 4.68 &$ -2.84 $& $ -26.27
\pm 0.37 $ \nl

26 & HD 123710 & 5750 & 4.55 & 0.75 & 6.98 &$ -0.54 $& $ 13.18 \pm
0.34 $ \nl

27 & HD 148816 & 5800 & 4.13 & 0.80 & 6.70 &$ -0.82 $&  $ -47.68
\pm 0.33 $ \nl

28 & HD 159482 & 5770 & 4.31 & 0.83 & 6.67 &$ -0.85 $& $ -139.94
\pm 0.36 $   \nl

29 & HD 165908   & 5970 & 4.24 & 1.18 & 6.89 &$ -0.63 $& $ -1.50
\pm 0.33 $ \nl

30 & HD 175305 & 5030 & 2.52    & 1.48 & 6.06 &$ -1.46 $& $
-184.60 \pm 0.40 $ \nl

31 & HD 201889 & 5700 & 4.33 & 0.95 & 6.71 &$ -0.81 $& $ -103.22
\pm 0.40 $ \nl

32 & HD 204543 & 4720 & 1.27 & 2.00 & 5.73 &$ -1.79 $& $ -98.74
\pm 0.39 $ \nl

33 & BD+22 4454 & 5150 & 4.43 & 0.00 & 6.98 &$ -0.54 $& $ -106.92
\pm 0.35 $ \nl

34 & BD+17 4708 & 6060 & 3.83 & 0.97 & 5.80 &$ -1.72 $& $ -296.70
\pm 0.49 $ \nl

35 & HD 221170 & 4580 & 1.32 & 2.00 & 5.42 &$ -2.10 $& $ -121.73
\pm 0.43 $ 
\enddata
\label{table:atpar}
\end{deluxetable}

\end{document}